\documentclass[12, a4paper]{article}
\usepackage[margin=1 in]{geometry}
\usepackage{color}
\usepackage[utf8]{inputenc}
\usepackage{amsmath}
\usepackage{tabularx,ragged2e,booktabs,caption}

\usepackage[labelfont=bf]{caption}
\usepackage[font=footnotesize]{caption}
\usepackage{amssymb,latexsym}
\usepackage{graphicx}
\usepackage{epstopdf}
\usepackage{booktabs}
\usepackage{amsfonts}
\usepackage{tikz}
\usepackage{array}
\usepackage{mathtools}
\usepackage{eucal}
\usepackage[T1]{fontenc}
\usepackage{float}
\usepackage{subfig}
\numberwithin{equation}{section}
\usepackage{tabularx,ragged2e,booktabs,caption}
\usetikzlibrary{shapes.geometric, arrows}
\usepackage{hyperref}

\usetikzlibrary{decorations.pathreplacing,angles,quotes}
\tikzstyle{startstop} = [rectangle, rounded corners, minimum width=4cm, minimum height=2cm,text centered, draw=black, fill=white!30]
\tikzstyle{arrow} = [thick,->,>=stealth]

\usetikzlibrary{positioning}

\begin{document}
	\title{\textbf{	Application of ensemble transform data assimilation methods for parameter estimation in reservoir modelling}}
	\author{\small{\textbf{S Ruchi and S Dubinkina}} \\ \small{CWI, P.O. Box 94079, 1098 XG Amsterdam, The Netherlands} \\ \small{E-mail: s.ruchi@cwi.nl}}
	
	\date{}
	\maketitle
	
	\begin{abstract}
		Over the years data assimilation methods have been developed to obtain estimations of uncertain model parameters by taking into account a few observations of a model state. The most reliable methods of MCMC are computationally expensive.   
		Sequential ensemble methods such as ensemble Kalman filers and particle filters provide a favourable alternative.
		However, Ensemble Kalman Filter has an assumption of Gaussianity. Ensemble Transform Particle Filter does not have this assumption and has proven to be highly beneficial for an initial condition estimation and a small number of parameter estimation in chaotic dynamical systems with non-Gaussian distributions. In this paper we employ Ensemble Transform Particle Filter (ETPF) and Ensemble Transform Kalman Filter (ETKF) for parameter estimation in nonlinear problems with 1, 5, and 2500 uncertain parameters and compare them to importance sampling (IS). The large number of uncertain parameters is of a particular interest for subsurface reservoir modelling as it allows to parametrise permeability on the grid. 
		We prove that the updated parameters obtained by ETPF lie within the range of an initial ensemble, which is not the case for ETKF. We examine the performance of ETPF and ETKF in a twin experiment setup, where observations of pressure are synthetically created based on the known values of parameters. 	 
		For a small number of uncertain parameters (1 and 5) ETPF performs comparably to ETKF in terms of the mean estimation.
		For a large number of uncertain parameters (2500) ETKF is robust with respect to the initial ensemble while ETPF is 
		sensitive due to sampling error. 
		Moreover, for the high-dimensional test problem ETPF gives an increase in the root mean square error after data assimilation is performed. 
		This is resolved by applying distance-based localization, 
		which however deteriorates a posterior estimation of the leading mode by largely increasing the variance
		due to a combination of less varying localized weights, not keeping the imposed bounds on the modes via the Karhunen-Loeve 
		expansion, and the main variability explained by the leading mode. A possible remedy is instead of applying localization to use only leading modes 
		that are well estimated by ETPF, which demands a knowledge at which mode to truncate. 
	\end{abstract}

	\maketitle
	
	\section{Introduction}
		
An accurate estimation of subsurface geological properties like permeability, porosity etc. is essential for many fields specially where such predictions can have large economic or environmental impact, for instance prediction of oil or gas reservoir locations. Knowing the geological parameters a so-called forward model is solved for the model state and a prediction can be made. The subsurface reservoirs, however, are buried thousands of feet below the earth surface and exhibit a highly heterogeneous structure, which makes it difficult to obtain their geological parameters. Usually a prior information about the parameters is given, which still needs to be corrected by observations of pressure and production rates. These observations  are, however,  known only at well locations that are often hundreds of meter apart and corrupted by errors. This gives instead of  a well-posed forward problem an ill-posed inverse problem of estimating uncertain parameters, since many possible combinations of parameters can result in equally good matches to the observations. 

Different inverse problem approaches for groundwater and petroleum reservoir modelling, generally termed as history matching, have been developed over the past years, e.g. in \cite{Oletal97} the authors implemented Markov chain Monte Carlo methods with different perturbations and tested it on a 2-D reservoir model; \cite{Reetal96} obtained reservoir parameter estimations using Gauss-Newton method; \cite{Veetal06} used Levenberg--Marquardt method to characterize reservoir pore pressure and permeability. A review of history matching developments has been written in the review paper \cite{OlCh11}.

For reservoir models the term data assimilation and history matching are used interchangeably, as the goal of  data assimilation is the same as that of history matching, where observations are used to improve  a solution of a model. Ensemble data assimilation methods such as Ensemble Kalman filters~\cite{Evensen09} have been originally developed in meteorology and oceanography for the state estimation. Now it is one of the frequently employed approaches for parameter estimation in subsurface flow models as well e.g. \cite{Oletal08}. A detailed review of ensemble Kalman filter developments in reservoir engineering can be found in \cite{Aaetal09}.  An ensemble Kalman filter efficiently approximates a true posterior distribution if the distribution is not far from Gaussian, as it corrects only the mean and the variance. For nonlinear models with multimodal distributions, however, an ensemble Kalman filter fails to correctly estimate the posterior, as shown in \cite{DoDR11}.

Importance Sampling (IS) is quite promising for such models as it does not have any assumptions of Gaussianity. 
It is also an ensemble based method in which the probability density function is represented by a number of samples. 
One sample corresponds to one configuration of uncertain model parameters. 
The forward model is solved for each sample and predicted data is computed. 
The weight is assigned to samples based on the observations of the true physical system and the predicted data. 
The drawback of IS is that it does not update the uncertain parameters but only their weight,
thus a computationally unaffordable ensemble is required. In order to decrease this cost
a family of particle filters~\cite{Doucet01} has been developed where IS is supplied
with resampling, and a sample is called particle. 
A significant work for parameter estimation using particle filtering has been done in hydrology. In \cite{Moetal05} authors used it to estimate model parameters and state posterior distributions for a rainfall-runoff model. \cite{WeEl06} compared an ensemble Kalman filter and a particle filter with different resampling strategies for a rainfall-runoff forecast and obtained that as the number of particles increases the particle filter outperforms the ensemble Kalman filter. \cite{Guietal12} employed particle filtering to correct the soil moisture and to estimate hydraulic parameters. 

The resampling in particle filtering is, however, stochastic. Ensemble Transform Particle Filter (ETPF) \cite{ReCo15} 
is a particle filtering method that {\it deterministically} resamples the particles based on their weights and covariance maximization among the particles. ETPF has been used for initial condition estimations and for parameter estimations in chaotic dynamical systems with a small number of uncertain parameters (Lorenz 63 model). It has not been applied, however, in subsurface reservoir modelling for estimating a large number of uncertain parameters. In this paper we employ it for estimating uncertain parameters in subsurface reservoir modelling. ETPF provides the equations that are solved in the space defined by the ensemble members. Therefore for comparison we employ Ensemble Transform Kalman Filter (ETKF) \cite{Bietal01} that also transforms the state from the model space to the ensemble space, minimises the uncertainty in the ensemble space and transforms the estimation back to the model space.  

In this paper we investigate the performance of ETPF and ETKF for parameter estimation in nonlinear problems and compare them to IS with a large ensemble. This paper is organized as follows: in Sect.~\ref{Sec:DA} we describe IS, ETPF, and ETKF for parameter estimation. We apply these methods in Sect.~\ref{Sec:1D} to a one parameter nonlinear test case, where the posterior can be computed analytically, and in Sect.~\ref{Sec:Darcy} to a single-phase Darcy flow, where the number of parameters is 5 and 2500. In Sect.~\ref{Sec:Con} we draw the conclusions.

  \section{Data assimilation methods} \label{Sec:DA}
 We implement an ensemble transform Kalman filter and an ensemble transform particle filter for estimating parameters of subsurface flow. Both of these methods are based on Bayesian framework. Assume we have an ensemble of $M$ model parameters $\{\vec{u}_m\}_{m=1}^M$, then  according to this framework, the posterior distribution, which is the probability distribution $\pi(\vec{u}_m|\vec{y}_{\text{obs}})$ of the model parameters $\vec{u}_m$  given a set of observations $\vec{y}_{\text{obs}}$, can be estimated by the pointwise multiplication of the prior probability distribution $\pi(\vec{u}_m)$ of the model parameters $\vec{u}_m$ and the conditional probability distribution $\pi(\vec{y}_{\text{obs}}|\vec{u}_m)$ of the observations given the model parameters, which is also referred as the likelihood function,
\begin{flalign*} \label{Bayes}
\pi(\vec{u}_m|\vec{y}_{\text{obs}}) = \frac {\pi(\vec{y}_{\text{obs}}|\vec{u}_m) \pi(\vec{u}_m)}{\pi(\vec{y}_{\text{obs}})}  \nonumber.
\end{flalign*}  
The denominator $\pi(\vec{y}_{\text{obs}})$ represents the marginal of observations and can be expressed as:
\begin{equation}
\pi(\vec{y}_{\text{obs}}) = \sum_{m=1}^{M}\pi(\vec{y}_{\text{obs}},\vec{u}_m) = \sum_{m=1}^{M}  \pi(\vec{y}_{\text{obs}}| \vec{u}_m) \pi(\vec{u}_m), \nonumber \end{equation}
which shows that $\pi(\vec{y}_{\text{obs}})$ is just a normalisation factor.

 \subsection{Ensemble Transform Kalman Filter} 
Assume we have initially an ensemble of $M$ model parameters $\{\vec{u}_m^b\}_{m=1}^{M}$, where $b$ refers to a background (prior) ensemble, which are sampled from a chosen prior probability density function, then the ensemble Kalman  estimate (or analysis)  $\{\vec{u}_m^a\}_{m=1}^{M}$ is given by:
\begin{equation*} 
{\vec{u}}^a_m  = \sum_{l=1}^{M} \text{diag}\left( s_{lm} + q_l - \frac{1}{M}\right)\vec{u}^b_l,\quad m = 1,\hdots,M,
\end{equation*}
where $\text{diag}$ is a diagonal matrix, $s_{lm}$ is the $(l,m)$ entry of a matrix $\textbf{S}$
\begin{equation}\label{S} 
\textbf{S} = \left[ \textbf{I}+\frac{1}{M-1} (\textbf{A}^b)^T \textbf{R}^{-1} \textbf{A}^b\right]^{-1/2},      
\end{equation}
and $q_l$ is the $l$-th entry of a column $\vec{q}$
\begin{equation*}
{\vec{q}} = \frac{1}{M-1}\textbf{1}_M - \textbf{S}^2(\textbf{A}^b)^T \textbf{R}^{-1} (\bar{\vec{y}}^b - \vec{y}_{\text{obs}} ).
\end{equation*}
Here \textbf{I} is an identity matrix of size $M \times M$, $\textbf{1}_M$ is a vector of size $M$ with all ones,
$\bar{\vec{y}}^b $ is the mean of the predicted data defined by
\begin{equation*}
\bar{\vec{y}}^b=\frac{1}{M}\sum_{m=1}^{M}\vec{y}_m^b,
\end{equation*}
$\textbf{A}^b$  is the background ensemble anomalies of the predicted data defined as
\begin{equation*}
\textbf{A}^b=  \begin{bmatrix}
(\vec{y}_1^b-\bar{\vec{y}}^b)      & (\vec{y}_2^b-\bar{\vec{y}}^b)  &\dots &(\vec{y}_M^b-\bar{\vec{y}}^b) 
\end{bmatrix}, 
\end{equation*}
and  $\textbf{R}$ is the measurement error covariance. To ensure that the anomalies of analysis remain zero centered we check whether $\textbf{A}^a\textbf{1}_M=\textbf{A}^b\textbf{S}\textbf{1}_M=\textbf{0}$, given $\textbf{S}\textbf{1}_M =\textbf{1}_M$ and $\textbf{A}^b\textbf{1}_M=\textbf{0}$. The model parameters $\vec{u}_m^b$ and the predicted data $\vec{y}_m^b$ are related by $\vec{y}_m^b = h(\vec{u}_m^b)$, where $h$ is a nonlinear function and here we assume that the function $h$ is known. 

\subsection{Ensemble Transform Particle Filter}\label{PF} 
In particle filtering  we represent the probability distribution function using ensemble members (also called particles) as in ensemble Kalman filter. We start by assigning prior (background) weights $\{w_m^b\}_{m=1}^M$ to $M$ particles and then compute new (analysis) weights  $\{w_m^a\}_{m=1}^M$ using the Bayes' formula and observations $\vec{y}_{\text{obs}}$
\begin{equation} \label{PF_bayes}
w_m^{a} = \frac {\pi (\vec{y}_{\text{obs}}| \vec{u}_m^b)w_m^{b} } {\pi (\vec{y}_{\text{obs}})}.
\end{equation}
We assume that initially all particles have equal weight, thus $w_m^b=1/M$ for $m=1,\hdots,M$, and that the likelihood is Gaussian with error covariance matrix \textbf{R}, then from Eq.~\eqref{PF_bayes} $w_m^a$ is given by
\begin{equation} \label{PF_wgt}
w^a_m = \frac{\text{exp}\left[-\frac{1}{2}(\vec{y}_m^b - \vec{y}_{\text{obs}})^T \textbf{R}^{-1} (\vec{y}_m^b - \vec{y}_{\text{obs}})\right]}{\sum_{j=1}^{M}\text{exp}\left[-\frac{1}{2}(\vec{y}_j^b - \vec{y}_{\text{obs}})^T \textbf{R}^{-1} (\vec{y}_j^b - \vec{y}_{\text{obs}})\right]},  \quad m = 1,\hdots,M.
\end{equation}
In Importance Sampling (IS), which will be used in this paper as a "ground" truth, these weights define the posterior pdf. The mean parameter for IS is then
\[
\bar{\vec{u}}^a = \sum_{m=1}^M\vec{u}_m^b w^a_m.
\]
It is important to note that IS \textit{does not} change the parameters $\vec{u}$, it only modifies the weight of the particles (samples). Therefore a resampling needs to be implemented for parameter estimation, which is usually stochastic. Instead particle filtering has been modified using a deterministic coupling methodology which resulted in an ensemble transform particle filter of~\cite{ReCo15}. ETPF looks for a coupling between two discrete random variables $B_1$ and $B_2$  so as to convert the ensemble members belonging to the random variable $B_2$ with probability distribution $\pi (B_2 = \vec{u}_m^b) =w^a_m  $ to the random variable $B_1$ with uniform probability distribution $\pi (B_1 = \vec{u}_m^b) = 1/M$. The coupling between these two random variables is an $M \times M$ matrix $\textbf{T}$ whose entries should satisfy 
\begin{align}  \label{Tconditions3}
t_{mj} & \geq 0,  \quad m,j = 1,\hdots,M, \\ \label{Tconditions1}
\sum_{m=1}^{M} t_{mj} & = \frac{1}{M},   \quad j = 1,\hdots,M,\\ \label{Tconditions2}
\sum_{j=1}^{M} t_{mj} & =w^a_m, \quad m = 1,\hdots,M.
\end{align} 
An optimal coupling matrix $\textbf{T}^{*}$  with elements $t_{mj}^{*}$ minimizes the squared Euclidean distance 
\begin{equation}\label{T}
J(t_{mj}) = \sum_{m,j=1}^{M} t_{mj} ||\vec{u}_m^b-\vec{u}_j^b ||^2 
\end{equation}
and the analysis model parameters are obtained by the linear transformation 
\begin{equation} \label{Zana}
\vec{u}_j^a = M\sum_{m=1}^{M} t_{mj}^*\vec{u}_m^b ,   \quad j = 1,\hdots,M.
\end{equation}
Then the mean parameter for ETPF is 
\[
\bar{\vec{u}}^a = \sum_{m=1}^M\vec{u}_m^a \frac{1}{M}.
\]
We use $FastEMD$ algorithm developed by Pele $\&$ Werman \cite{PeWe09} to solve the linear transport problem and get the optimal transport matrix. 

\textbf{Remark:} An important property of ETPF is preservation of imposed  interval bounds on ensemble members. Consider an ensemble of parameters $\{\vec{u}_m^b\}_{m=1}^M$  given by
\begin{equation*}
\vec{u}_m^b = (a_m^b  \ b_m^b \ c_m^b)^T,  \quad m = 1,\hdots,M, 
\end{equation*}
where we assume all the parameters $\{a_m^b\}_{m=1}^M$, $\{b_m^b\}_{m=1}^M$ and $\{c_m^b\}_{m=1}^M$ are bounded between $0$ and $1$. Therefore, the following inequalities hold:
\begin{align*} 
& 0 < a_\textrm{min} \leq a_m^b \leq a_\textrm{max} <1, \quad m=1,\hdots,M, \\
& 0 < b_\textrm{min} \leq b_m^b \leq b_\textrm{max} <1,  \quad m=1,\hdots,M, \\
& 0 < c_\textrm{min} \leq c_m^b \leq c_\textrm{max} <1, \quad m=1,\hdots,M. 
\end{align*} 
Now we assume two discrete random variables $B_1$ and $B_2$ have probability distributions given by
\begin{equation}
\pi (B_1 = \vec{u}_m^b) = 1/M, \quad \pi (B_2 = \vec{u}_m^b) =w_m^a, \nonumber
\end{equation}
with $w_m^a \geq 0$, $m=1,\hdots ,M$ and $\sum_{m=1}^{M}w_m^a =1 $. As ETPF looks for a matrix $\textbf{T}^{*}$ which defines coupling between these two probability distributions, each entry of this coupling matrix satisfies the conditions given by Eq.~\eqref{Tconditions3}--\eqref{Tconditions2}. These conditions assure that each entry of the coupling matrix will be non-negative and less than 1.
Since the  analysis given  by Eq.~\eqref{Zana} is
\begin{equation}
\vec{u}^a_m = \begin{bmatrix}
a_1^b(Mt^{*}_{1m})+a_2^b(Mt^{*}_{2m})+ \dots +a_M^b(Mt^{*}_{Mm})\\
b_1^b(Mt^{*}_{1m})+b_2^b(Mt^{*}_{2m})+ \dots +b_M^b(Mt^{*}_{Mm})\\
c_1^b(Mt^{*}_{1m})+c_2^b(Mt^{*}_{2m})+ \dots +c_M^b(Mt^{*}_{Mm})\\
\end{bmatrix},  \quad m=1,\dots,M, \nonumber
\end{equation}
these conditions lead to 
\begin{align*}
& 0 < a_\textrm{min} \leq a_m^a \leq a_\textrm{max} <1, \quad m=1,\dots,M, \\
& 0 < b_\textrm{min} \leq b_m^a \leq b_\textrm{max} <1,  \quad m=1,\dots,M, \\
& 0 < c_\textrm{min} \leq c_m^a \leq c_\textrm{max} <1,  \quad m=1,\dots,M.  
\end{align*}
Thus the coupling matrix bounds the analysis ensemble members to be in the desired range. This is not observed in ETKF as the matrix $\textbf{S}$ given by Eq.~\eqref{S} does not impose any of the non-equality and equality constraints,  so it results in values outside the bound.

\subsection{Localization}  
All variations of ensemble Kalman filter and particle filter are limited by the ensemble size. 
Since, even if the dimension of the problem is just up to a few thousands, a large ensemble size will make each run of the 
model computationally very expensive. This limit of a small ensemble size introduces sampling errors.  
To deal with this issue localized ETKF (LETKF) was introduced in \cite{Huetal07} and localized ETPF (LETPF) in \cite{ReCo15}. More recent approaches to particle filter localization include~\cite{PeMi16} and~\cite{Po16}.

For the local update of a model parameter $\vec{u}_m(X_i)$ at a grid point $X_i$, we  introduce a diagonal matrix $\hat{\textbf{C}}_i \in R^{N_y \times N_y}$ in the observation space with an element 
\begin{equation}\label{rho_matrix}
(\hat{\textbf{C}}_i)_{ll} =\rho \left(\frac{||{{X}_i -{r}_l}||}{r_\textrm{loc}}\right),
\end{equation}
where $i=1,\hdots,n^2$,   $l=1,\hdots,N_y$, $n^2$ is the number of model parameters, $N_y$ is the dimension of the observation space, ${r}_l$ denotes the location of the observation, $r_\textrm{loc}$ is a localisation radius and $\rho(\cdot)$ is a taper function, such as Gaspari-Cohn function \cite{GaCo99}
\begin{equation}
\rho(r) = 
\begin{cases}
1-\frac{5}{3}r^2 + \frac{5}{8} r^3+\frac{1}{2}r^4-\frac{1}{4}r^5, &  \quad 0 \leq r \leq 1, \\
-\frac{2}{3}r^{-1}+4-5r+\frac{5}{3}r^2+\frac{5}{8}r^3-\frac{1}{2}r^4+\frac{1}{12}r^5, &  \quad 1 \leq r \leq 2, \\
0, & \quad 2 \leq r.
\end{cases}  \nonumber
\end{equation}
Then the estimated model parameter at the location $X_i$ is 
\begin{equation*}
{\vec{u}}^a_m(X_i)  = \sum_{l=1}^{M} \text{diag}\left( s_{lm}(X_i) + q_l (X_i)- \frac{1}{M}\right)\vec{u}^b_l(X_i),\quad m = 1,\hdots,M,
\end{equation*}
where $\text{diag}$ is a diagonal matrix, $s_{lm}(X_i)$ is the $(l,m)$ entry of  the localized transformation matrix $\textbf{S}(X_i)$
\begin{equation*}
\textbf{S}(X_i) = \left[ \textbf{I}+\frac{1}{M-1} (\textbf{A}^b)^T (\hat{\textbf{C}}_i\textbf{R}^{-1}) \textbf{A}^b\right]^{-1/2}
\end{equation*}
and $q_l(X_i)$ is the $l$-th entry of the localized column $\vec{q}(X_i)$
\begin{equation*}
\vec{q}(X_i) = \frac{1}{M-1}\textbf{1}_M - \textbf{S}(X_i)^2(\textbf{A}^b)^T \textbf{R}^{-1} (\bar{\vec{y}}^b - \vec{y}_{\text{obs}} ).
\end{equation*}

LETPF modifies the likelihood and thus the weights given by Eq.~\eqref{PF_wgt} are computed locally at each grid $X_i$
\begin{equation}\label{PF_wgtR0}
w^a_m({X}_i) = \frac{\text{exp}\left[-\frac{1}{2}(\vec{y}_m^b - \vec{y}_{\text{obs}})^T ( \hat{\textbf{C}}_i\textbf{R}^{-1}) (\vec{y}_m^b - \vec{y}_{\text{obs}})\right]}{\sum_{j=1}^{M}\text{exp}\left[-\frac{1}{2}(\vec{y}_j^b - \vec{y}_{\text{obs}})^T ( \hat{\textbf{C}}_i\textbf{R}^{-1}) (\vec{y}_j^b - \vec{y}_{\text{obs}})\right]}, \quad m=1,\hdots,M, 
\end{equation}
where $\hat{\textbf{C}}_i$ is the diagonal matrix given by Eq.~\eqref{rho_matrix}. Then the estimated model parameter $ \vec{u}_j^a(X_i)$ at the grid $X_i$ is given by
\begin{equation*}
\vec{u}_j^a(X_i) = M\sum_{m=1}^{M} t_{mj}^*\vec{u}(X_i)_m^b ,   \quad j = 1,\hdots,M,
\end{equation*}
where $t_{mj}^*$ is an element of an optimal coupling matrix $\textbf{T}^{*}$  which minimizes the squared Euclidean distance at the grid point $X_i$
\begin{equation}\label{TR0}
J(t_{mj}) = \sum_{m,j=1}^{M} t_{mj}  [u_{m}^b(X_i)-u_{j}^b(X_i)]^2,
\end{equation}
which reduces LETPF to a univariate transport problem.
It should be noted that localization can be applied only for grid-dependent parameters.
 
  \section{One parameter nonlinear problem} \label{Sec:1D}
\begin{figure}[t]
	\centering
	\includegraphics[width=1.0\linewidth]{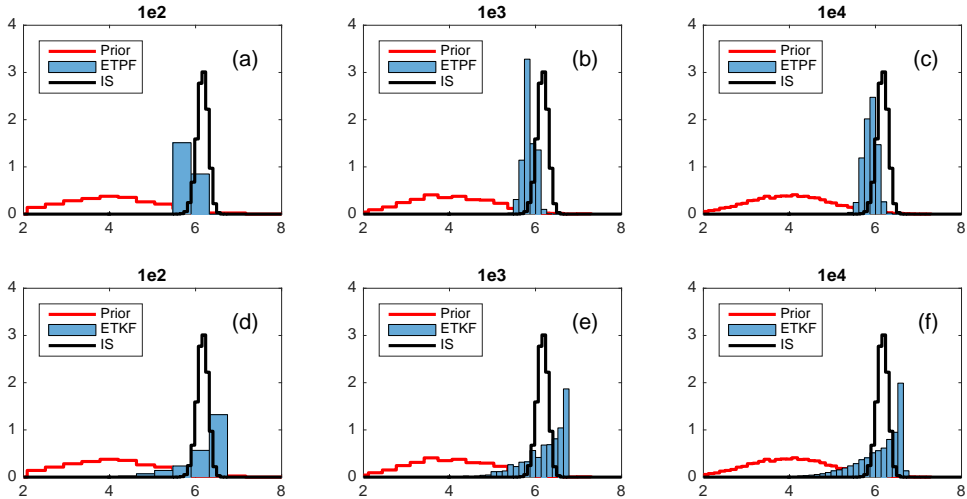}
	\caption{Probability density functions for the one parameter nonlinear problem. Top: ETPF, bottom: ETKF. Left: ensemble size $10^2$, center: ensemble size $10^3$, right: ensemble size $10^4$. Prior is in red. True pdf obtained by IS with ensemble size $10^5$ is in black.}
	\label{Fig_1Dpdf}
\end{figure}

First we consider a one parameter nonlinear problem from~\cite{ChOl13}. 
The prior distribution is Gaussian distribution with mean 4 and variance 1.  The nonlinear forward model is  
\[
h(u) = \frac{7}{12} u^3 -\frac{7}{2}u^2 + 8u.
\]
The true parameter $u^{\rm true}$ gives $h(u^{\rm true}) = 48$ and the observation error is drawn from a Gaussian distribution with zero mean and variance 16. In Fig.~\ref{Fig_1Dpdf} we plot the posterior probability density functions estimated by ETPF (top), ETKF (bottom) with ensemble sizes $10^2$ (left), $10^3$ (center), and $10^4$ (right). The prior distribution is shown in red and the posterior estimated by IS with ensemble size $10^5$ is shown in black. We can see that ETPF provides better approximation of the true probability density function, while ETKF gives a skewed posterior. It should be noted that ETKF is able to give a non-Gaussian (though wrong) posterior due to the nonlinearity of the map between the uncertain parameters and observations. 

\section{Single-phase Darcy flow} \label{Sec:Darcy}

We consider a steady-state single-phase Darcy flow model defined over an aquifer of two-dimensional physical domain  $D = [0,1] \times [0,1] $, which is given by,
\begin{align*} 
-\nabla \cdot (k(x,y)\nabla P(x,y)) & =f(x,y),   \quad  (x,y)\in D  \\
P(x,y) & =0, \quad  (x,y) \in \partial D 
\end{align*} 
where $\nabla = ( \partial / \partial x \ \partial / \partial y)^T$, $\cdot$ denotes the dot product, $P(x,y)$  the pressure, $k(x,y)$ the permeability, $f(x,y)$  the source term, which we assume to be $2\pi^2 \text{cos}(\pi x) \text{cos}(\pi y)$, and $\partial D$  the boundary of domain $D$.  The forward problem of this second order elliptical equation is to find the solution of pressure $P(x,y)$ for given $f(x,y)$ and $k(x,y)$. We, however, are interested in finding permeability given noisy observations of pressure at a few locations. 

We perform numerical experiments with synthetic observations, where instead of a measuring device a model is used to obtain observations. We implement a cell-centered finite difference method to discretize the domain $D$ into $n\times n$ grid cells $X_i$ of size $\Delta x^2$ and solve the forward model with the true parameters. Then the synthetic observations are obtained by
\begin{equation*} 
\vec{y}_{\text{obs}} = \textbf{L}(\textbf{P}) + \eta,              
\end{equation*}
with an element of $\textbf{L}(\textbf{P})$ being a linear functional of pressure, namely
\begin{equation*}
L_l(\textbf{P}) = \frac{1}{2 \pi \sigma ^2} \sum_{i=1}^{n^2} \text{exp} \left(-\frac{||X_{i}-r_l ||^2}{2 \sigma^2}\right) P_{i} \Delta x^2, \quad l\in 1, \dots,N_y
\end{equation*}
where $n=50$, $\sigma = 0.01$, ${r}_l$ denotes the location of the observation and $N_y=16$, which is the number of observations. The observation locations are spread uniformly across the domain $D$ and $\eta$ denotes the observation noise drawn from a normal distribution with zero mean and standard deviation of $0.09$. This form of the observation functional and parametrization of the uncertain parameters given below guaranty the continuity of the forward map from the uncertain parameters to the observations and thus the existence of the posterior distribution as shown in \cite{Igetal14}. 

\subsection{Five parameter nonlinear problem}

\begin{figure} [t]
	\centering			
	\includegraphics[width=0.35\linewidth]{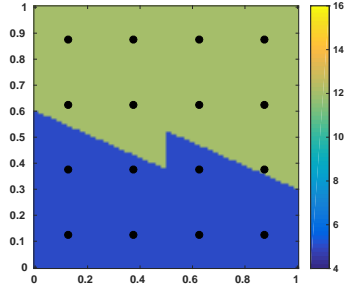}
	\caption{True permeability of the 5 parameter nonlinear problem with dots representing the observation locations.}\label{Fig_truth2L}
\end{figure}

For our first numerical experiment with Darcy flow, we consider a low-dimensional problem where the permeability field is defined by mere 5 parameters similarly to~\cite{Igetal14}. We assume that the entire domain $D=[0,1] \times [0,1]$ is divided into two subdomains $D_1$ and $D_2$ as shown in Fig.~\ref{Fig_truth2L}. Each subdomain of $D$ represents a layer and is assumed to have a  permeability function $k(\textbf{X})$, where an element of $\textbf{X}$ is defined by ${X_i}$ for $i=1,\hdots,n^2$. 
Parameters $a$ and $b$ denote the thickness of the bottom layer on either side, which correspondingly defines the slope of the interface. A parameter $c$ defines a vertical fault.The layer moves up or down depending on $c<0$ or $c>0$, respectively, and its location is assumed to be fixed at $x=0.5$.

Further, for this test case we assume piecewise constant permeability within each of the subdomains, hence  $k(\textbf{X})$ is given by
\begin{equation*} 
k(\textbf{X}) =  k_1 \delta _{D_1} (\textbf{X}) + k_2 \delta _{D_2} (\textbf{X}),
\end{equation*}
where $k_1$ and $k_2$ represent permeability of the subdomain $D_1$ and $D_2$, respectively, and $\delta$ is Dirac function. Then the parameters defining the permeability field for this configuration are    
\begin{equation*}
\vec{u} =(a \ b \ c \ \log(k_1) \ \log(k_2))^T. 
\end{equation*} 
We assume that the true parameters are $a^{\textrm{true}}=0.6$, $b^{\textrm{true}}=0.3$, $c^{\textrm{true}} = -0.15$, $k_1^{\textrm{true}} = 12$ and $k_2^{\textrm{true}} = 5$. These parameters are used to create synthetic observations. Figure~\ref{Fig_truth2L} shows the true permeability with dots representing the observation locations. Next, we assume that the five uncertain parameters are drawn from a uniform distribution over a specified interval, namely $a, b\sim\mathcal{U}[0,1]$, $c \sim \mathcal{U}[-0.5,0.5]$, $k_1 \sim \mathcal{U}[10,15]$ and $k_2 \sim \mathcal{U} [4,7]$.  

As it was pointed out in Sect.~\ref{PF}, ETPF updates the parameters within the original range of an initial ensemble, while ETKF does not. Therefore a change of variables has to be performed for ETKF so that the updated parameters are physically viable. In order to be consistent we perform the change of variables for ETPF as well. As the domain $D$ is $[0,1] \times [0,1]$, the parameters $a$ and $b$ should lie within the interval $[0,1]$. To enforce this constraint we substitute $a$ according to
\begin{equation*}
a^\prime =\text{log}\left(\frac{a}{1-a}\right), \quad a^\prime \in R
\end{equation*}
and similarly $b$ is substituted by $b^\prime$. Thus the uncertain parameters are now $\vec{u}^\prime = (a^\prime\  b^\prime\  c\ \log(k_1)\ \log(k_2))^T$. 

\begin{figure} [t]
	\centering		
	\includegraphics[width=1.0\linewidth]{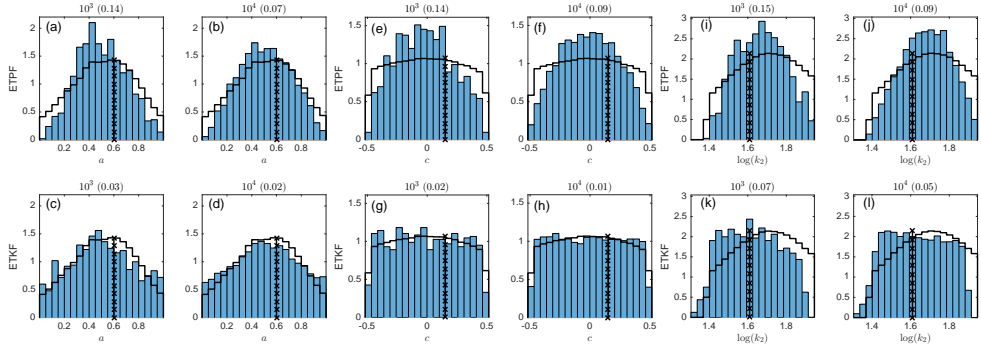}	
	\caption{Probability density functions for the parameters $a$ (a)--(d), $c$ (e)--(h), and $\log(k_2)$ (i)--(l). The posterior obtained by IS with ensemble size $10^6$ is plotted as a black line and the true values of parameters are plotted as  black crosses. The posterior of ETPF is shown at the top and the posterior of ETKF at the bottom. ETPF and ETKF used $10^3$ (odd columns) and $10^4$ (even columns) ensemble members. }	
	\label{Fig_5Dpdf}
\end{figure}

In Fig.~\ref{Fig_5Dpdf} we plot probability density functions for parameters $a$ (a)--(d), $c$ (e)--(h) and $\log(k_2)$ (i)--(l), 
as the parameters $b$ and $\log(k_1)$ show similar results. 
The posterior obtained by IS with ensemble size $10^6$ is plotted as a black line and the true value of parameters is plotted as a black line with crosses. 
The posterior of ETPF is shown at the top and the posterior of ETKF at the bottom. 
ETPF and ETKF used $10^3$ (odd columns) and $10^4$ (even columns) ensemble members. 
In order to perform an objective comparison between the probabilities we compute the Kullback-Leibler divergence of a 
posterior $\pi$ obtained by either ETPF or ETKF and the posterior $\pi^\textrm{IS}$ obtained by IS
\begin{equation}\label{DKL}
\textrm{D}_\textrm{KL} (\pi^\textrm{IS} \parallel \pi) = \sum_{i=1}^{N_\textrm{b}} \pi^\textrm{IS}(u_i) \log \frac{\pi^\textrm{IS}(u_i)}{\pi(u_i)}(u_i-u_{i-1}), 
\end{equation}
where $N_\textrm{b}=20$ is the number of bins. The Kullback-Leibler divergence for parameters $a$, $c$ and $\log(k_2)$
is displayed in the titles of Fig.~\ref{Fig_5Dpdf}, where we observe that ETKF outperforms ETPF.


\begin{figure} [t]
	\centering		
	\includegraphics[width=1.0\linewidth]{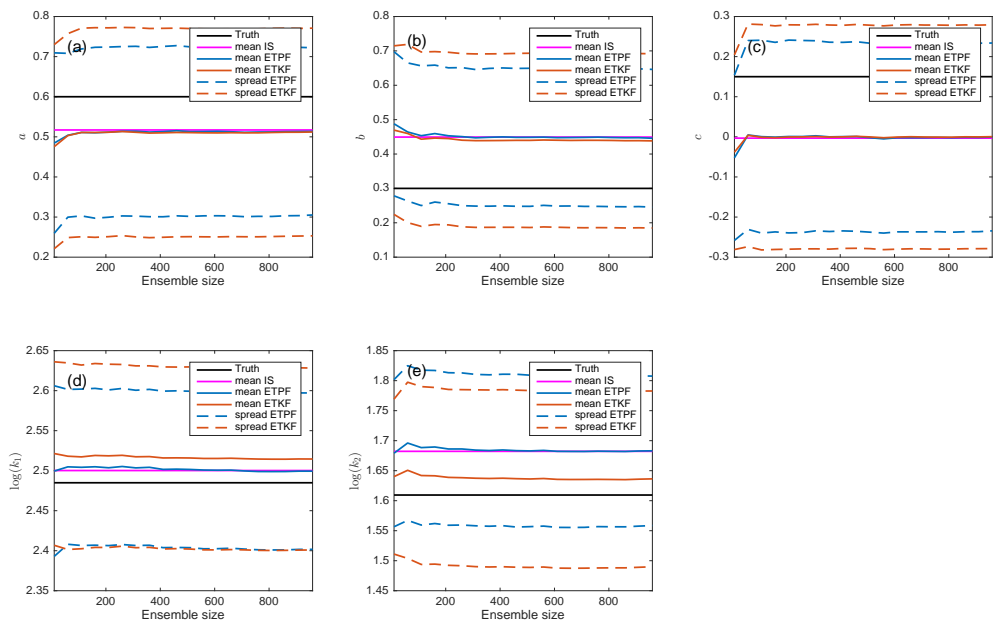}	
	\caption{${\bar{\bar{\vec{u}}}}^a$ and ${\bar{\bar{\vec{u}}}}^a \pm {\bar{\vec{u}}^a_\textrm{std}}$ w.r.t 
		ensemble size: (a) for the parameter $a$, (b) for $b$, (c) for $c$, (d) for $\log(k_1)$, (e) for $\log(k_2)$. 
		ETPF is shown in blue,  ETKF in red, the true parameters are in black and the mean of IS in magenta.}	
	\label{Fig_5Dpar}
\end{figure}

In order to check the sensitivity of the results to the initial parameter ensemble we perform $10$ simulations based on a random draw of an initial ensemble from the same prior distributions. We conduct the numerical experiments for ensemble sizes varying from $10$ to $10^3$ with an increment of $50$.  
In Fig.~\ref{Fig_5Dpar} we plot the true parameters, the mean estimated by IS, 
the mean ${\bar{\bar{\vec{u}}}}^a$ and the spread ${\bar{\bar{\vec{u}}}}^a \pm {\bar{\vec{u}}_\textrm{std}^a}$ 
of estimated parameters averaged over $10$ simulations
\begin{equation*}
{\bar{\bar{u}}}_i^a = \frac{1}{10}\sum_{r=1}^{10} {\bar{u}}_i^{a,r}, \quad
\bar{u}^a_{\textrm{std}}  =\frac{1}{10} \sum_{r=1}^{10}\sqrt[]{\frac{1}{M-1}\sum_{m=1}^{M} (u_{i,m}^{a,r}-\bar{u}_i^{a,r})^2},\ \mbox{where} \ \bar{u}_i^{a,r} =\frac{1}{M}\sum_{m=1}^{M} u_{i,m}^{a,r}, \ r = 1,\hdots,10,
\end{equation*}
$M$ is ensemble size, $i=1,\hdots,5$ is parameter index, and the superscript $a$ is for analysis. 
We observe that all the methods including IS have a bias in the estimations of geometrical
parameters, which is due to a small number of observations.
ETPF and ETKF perform comparably in terms of mean estimation, 
though some are better estimated by ETKF and other are better estimated by ETPF. 
Comparing the error in pressure of the mean parameters we observe that the methods 
are equivalent (thus not shown), which is a manifestation of the ill-posedness of the problem.
In Fig.~\ref{Fig_5Dpar} we see that the spread from ETPF is smaller than from ETKF for each parameter. 
Both methods are slightly underdispersive as the spread to error ratio is below 1. For ensemble size $10^3$ ETKF gives $(0.95\     0.88\    0.88\   0.97\    0.98)$ and ETPF gives $(0.92\    0.81\    0.84\    0.99\    0.86)$ 
for $(a\ b\ c\ \log(k_1)\ \log(k_2))$. Thus ETKF gives better ratio for all the parameters but $\log(k_1)$. 

\begin{figure} [t]
	\centering		
	\includegraphics[width=1.0\linewidth]{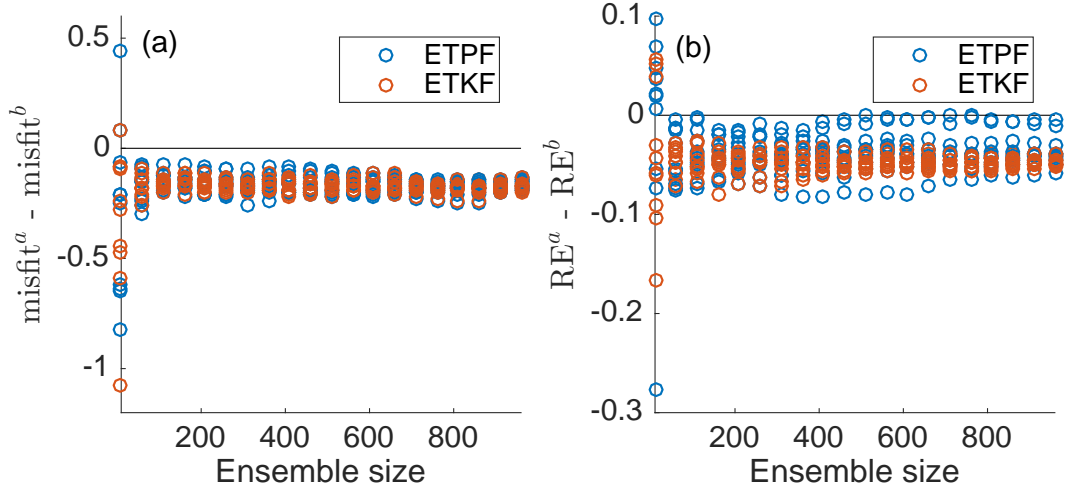}	
	\caption{$\text{misfit}^{a,r}- \text{misfit}^{b,r}$ (a) and $\text{RE}^{a,r}-\text{RE}^{b,r}$ (b) w.r.t ensemble size. ETPF is shown in blue, ETKF in red and the zero level in black. One circle is for one simulation.}	
	\label{Fig_5Dchange}
\end{figure}

We compute an average of the relative error over all parameters
\begin{equation*} 
\text{RE}^{a,r}  =\frac{1}{5}\sum_{i=1}^5
\frac{|\bar{u}_i^{a,r}-u_i^\textrm{true}|}{|u_i^\textrm{true}|}, \  r=1,\hdots,10,
\end{equation*} 
and the data misfit
\begin{equation}  \label{misfitar}
\text{misfit}^{a,r}  = (\bar{\vec{y}}^{a,r} - \vec{y}_\text{obs})^TR^{-1}(\bar{\vec{y}}^{a,r} - \vec{y}_\text{obs}), \  r=1,\hdots,10
\end{equation}
after data assimilation. The same metrics are computed before data assimilation and denoted by a superscript $b$. In Fig.~\ref{Fig_5Dchange}(a)--(b) we plot $(\text{misfit}^{a,r}- \text{misfit}^{b,r})$ and $(\text{RE}^{a,r}-\text{RE}^{b,r})$, respectively, for each simulation $r$ as a function of ensemble size. ETPF is shown in blue and ETKF in red. Black line is at zero level. Positive values of the differences mean an increase of either data mismatch or relative error after data assimilation. We observe a data misfit decrease for both ETPF and ETKF except at an ensemble size 10. RE does not always decrease for ETPF: for some simulations ETPF is at zero level or slightly above it, while for ETKF the sole exception is at an ensemble size 10.

\subsection{High-dimensional nonlinear problem}

Next, we consider a high-dimensional problem where the dimension of the uncertain parameter is $n^2 = 2500$. The domain $D$ is now not divided into subdomains. However, unlike in the previous test case  here we implement a spatially varying permeability field.
We assume the log permeability is generated by a random draw from a Gaussian distribution $\mathcal{N}(\log(\textbf{5}),\textbf{C})$. Here  $\textbf{5}$ is an  $n^2$ vector with all 5. $\textbf{C}$ is assumed to be an exponential correlation with an element of $\textbf{C}$ being 
\begin{equation*} 
{C}_{i,j}  = \text{exp}(-3(|h_{i,j}|/v)), \ i,j=1,\hdots,n^2.
\end{equation*}
Here $h_{i,j}$ is the distance between two spatial locations  and $v$ is the correlation range which is taken to be $0.5$.
For the log permeability we use Karhunen-Loeve expansions of the form
\begin{equation}\label{KLexp}
\log(k_j) = \log(5)+ \sum_{i=1}^{n^2} \sqrt{\lambda_{i}} \nu_{i,j}  \mathcal{Z}_{i}, \quad \mbox{for}  \quad j=1,\hdots,n^2,
\end{equation}
where $\lambda$ and $\nu$ are eigenvalues and eigenfunctions of $\textbf{C}$, respectively,  and the vector $\mathcal{Z}$ is of dimension  $n^2$ iid from a Gaussian distribution with zero mean and variance one.
Making sure that the eigenvalues are sorted in descending order $\mathcal{Z}_{i}  \sim \mathcal{N}(0,1)$ produces 
$\log(\textbf{k}) \sim \mathcal{N}(\log(\textbf{5}),\textbf{C})$.
The uncertain parameter is thus $\vec{u} =  \mathcal{Z}$ with the dimension $n^2 = 2500$.

\begin{figure} [t]
	\centering		
	\includegraphics[width=1.0\linewidth]{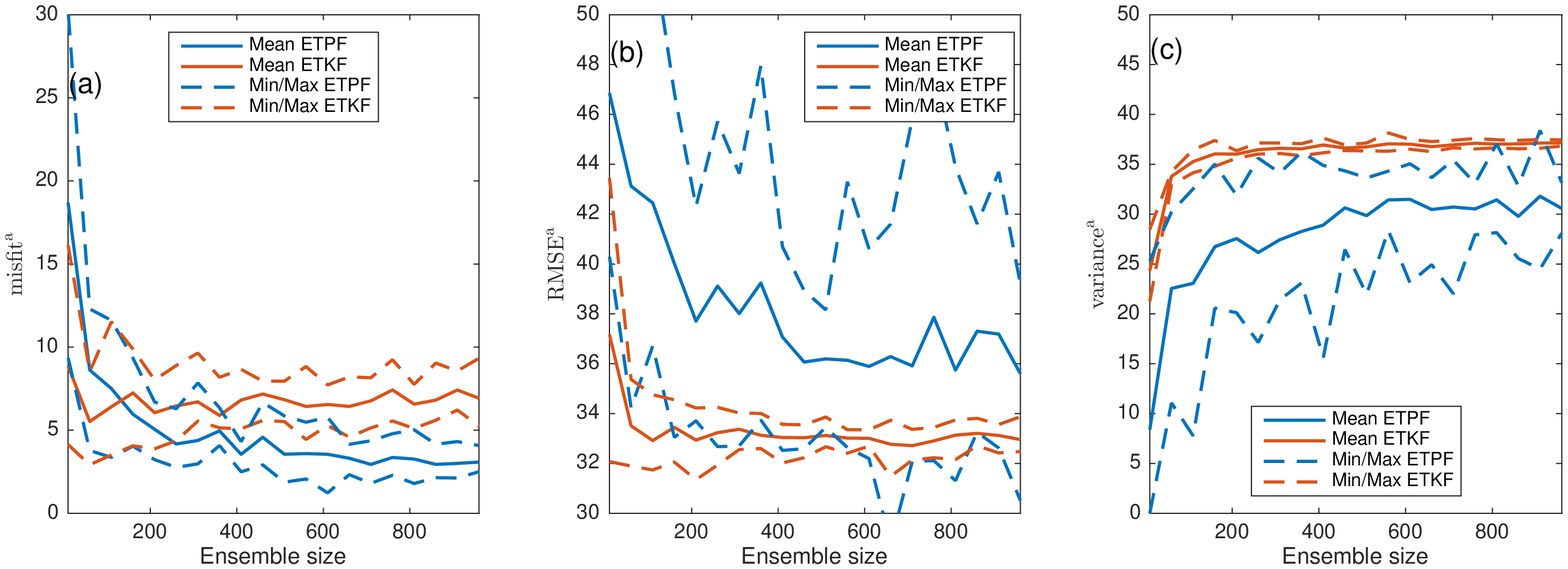}	
	\caption{Mean, minimum and maximum over 10 simulations after data assimilation for the data misfit (a), RMSE (b), and variance (c). ETPF is shown in blue and ETKF in red.}	
	\label{Fig_HDerr}
\end{figure}

\begin{figure} [t]
	\centering		
	\includegraphics[width=1.0\linewidth]{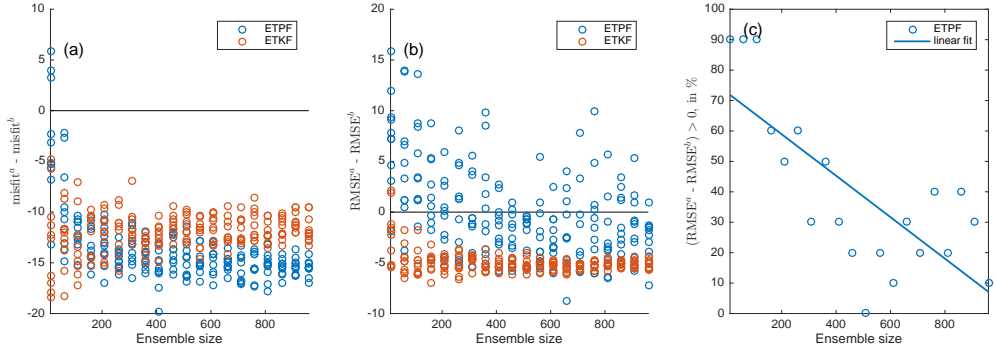}	
	\caption{$\text{misfit}^{a,r}- \text{misfit}^{b,r}$ (a) and $\text{RMSE}^{a,r}-\text{RMSE}^{b,r}$ (b) w.r.t ensemble size. ETPF is shown in blue, ETKF in red and zero level in black. One circle is for one simulation. For ETPF $\%$ of simulations that result in $(\text{RMSE}^{a}-\text{RMSE}^{b})>0$ and a linear fit as a function of ensemble size are shown in (c).}	
	\label{Fig_HDchange}
\end{figure}

We perform 10 different simulations based on a random draw of an initial ensemble from the prior distribution. We conduct the numerical experiments for ensemble sizes varying from $10$ to $10^3$ with an increment of $50$. 
We compute the root mean square error (RMSE) of the log permeability field 
\begin{equation*}
\text{RMSE}^{r,a} =\sqrt[]{\left(\log( \overline{\textbf{k}}^{a,r})-\log(\textbf{k}^\textrm{true})\right)^T \left( \log(\overline{\textbf{k}}^{a,r})-\log(\textbf{k}^\textrm{true})\right)} ,  \quad r=1,\hdots,10, \nonumber
\end{equation*}
and variance 
\begin{equation*}
\text{variance}^{r,a} = \frac{1}{M-1}\sum_{m=1}^M \left((\log(\textbf{k}^{a,r}_m) - \log(\overline{\textbf{k}}^{a,r})\right)^T \left(\log(\textbf{k}^{a,r}_m) -  \log(\overline{\textbf{k}}^{a,r})\right),  \quad r=1,\hdots,10. \nonumber
\end{equation*}
We also compute the data misfit for each simulation after data assimilation by Eq.~\eqref{misfitar}. 
In Fig.~\ref{Fig_HDerr} we plot mean, minimum and maximum over 10 simulations after data assimilation for the data misfit (left), RMSE (center), and variance (right). ETPF is shown in blue and ETKF in red. We observe that ETPF is underdispersive compared to ETKF as particle filters are highly degenerative compared to Kalman filters. Misfit given by ETPF is smaller than the one given by ETKF for almost all simulations at ensemble sizes greater than 150. The RMSE on the contrary is larger.
In Fig.~\ref{Fig_HDchange}(a)--(b) we plot $(\text{misfit}^{a,r}- \text{misfit}^{b,r})$ and $(\text{RMSE}^{a,r}-\text{RMSE}^{b,r})$, respectively, as a function of ensemble size for a simulation $r=1,\hdots,10$. The superscript $b$ is for the metrics before data assimilation and the superscript $a$ is for the metrics after data assimilation. ETKF always provides a decrease in both the data misfit and RMSE except at ensemble size 10. ETPF gives a decrease in the data misfit though an increase in RMSE, which indicates that ETPF overfitts the data. However, as the ensemble size increases this happens less often as can be seen in Fig.~\ref{Fig_HDchange}(c), where we plot for ETPF a percentage of simulations that result in $(\text{RMSE}^{a}-\text{RMSE}^{b})>0$ and a linear fit as a function of ensemble size.

\begin{figure} [t]
	\centering		
	\includegraphics[width=1.0\linewidth]{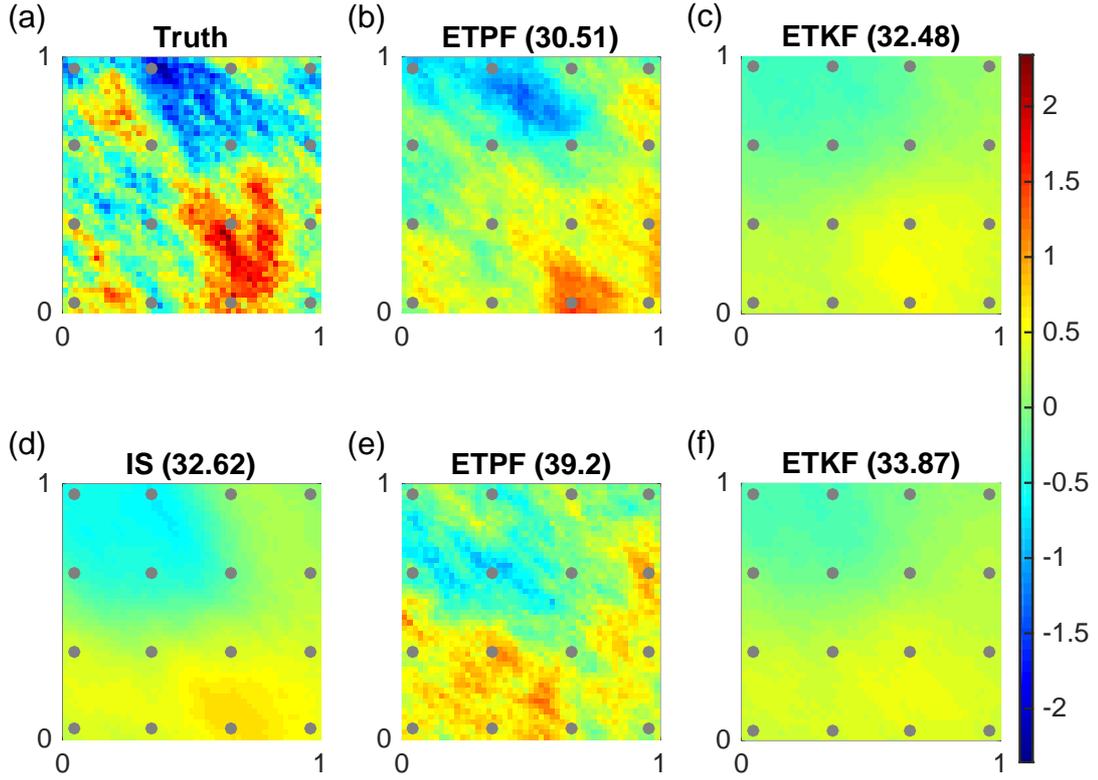}	
	\caption{Log permeability field with dots representing the observation locations. Truth is shown in (a) and mean obtained by IS with ensemble size $10^5$ in (d).
		Mean obtained with ensemble size $10^3$ by ETPF shown in (b--e) and by ETKF in (c--f),
		where (b--c) are at the smallest RMSE and (e--f) are at the largest RMSE over simulations. The corresponding RMSE is given in brackets.}
	\label{Fig_HDMF}
\end{figure}

\begin{figure} [t]
	\centering		
	\includegraphics[width=1.0\linewidth]{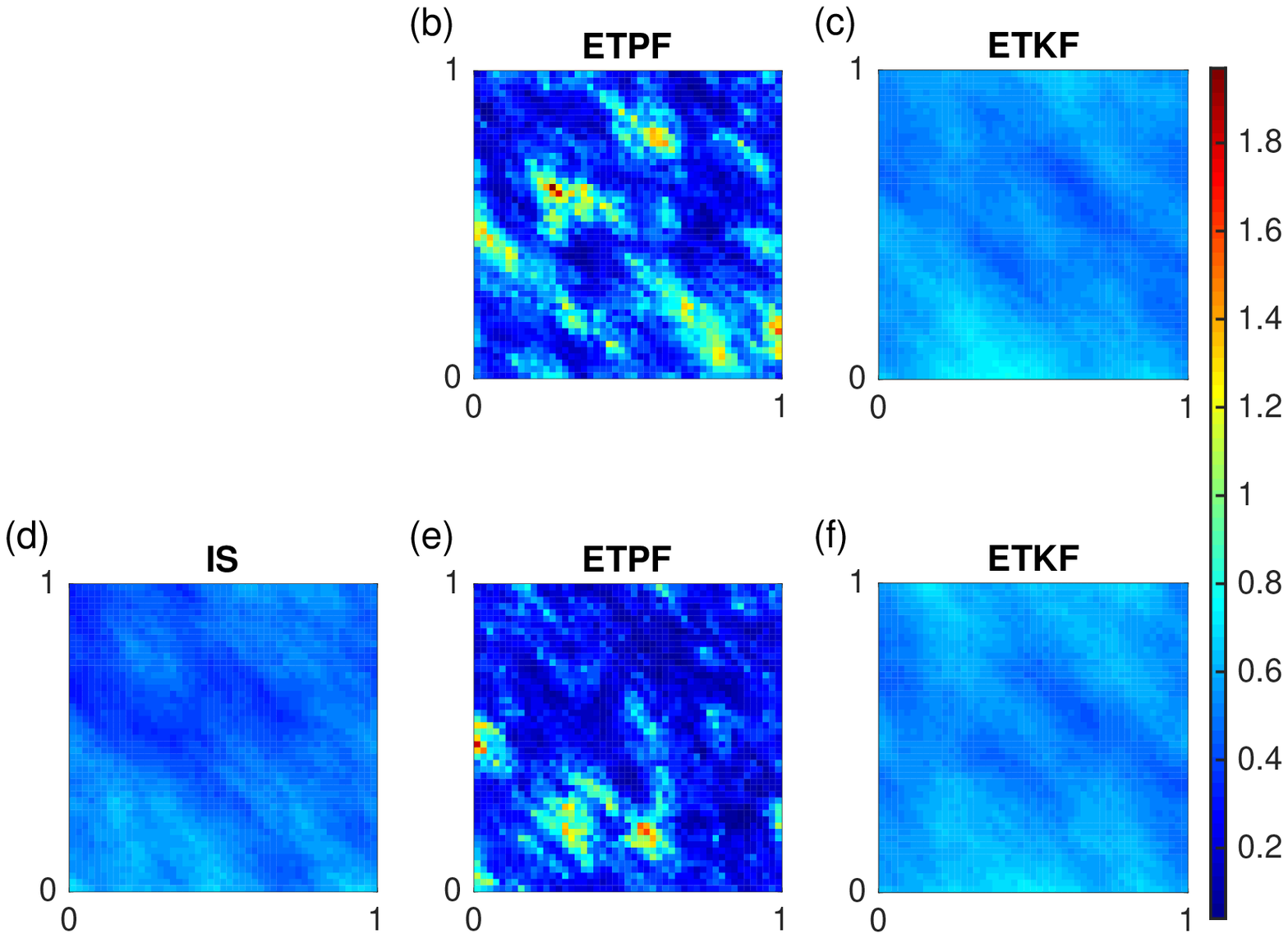}	
	\caption{Variance of log permeability fields: obtained with ensemble size $10^5$ by IS (d), with ensemble size $10^3$ by ETPF (b--e), and ETKF (c--f). Variance at the smallest RMSE (b--c) and at the largest RMSE (e--f) over simulations.}	
	\label{Fig_HDVF}
\end{figure}

In Fig.~\ref{Fig_HDMF} we plot log permeability fields. In Fig.~\ref{Fig_HDMF}(a) the true permeability is shown with dots representing the observation locations, and in Fig.~\ref{Fig_HDMF}(d) the mean permeability field obtained by IS with ensemble size $10^5$. The RMSE provided by IS is 32.62.
In Fig.~\ref{Fig_HDMF}(b--e) and Fig.~\ref{Fig_HDMF}(c--f) we display mean permeability fields obtained with ensemble size $10^3$ by ETPF and ETKF, respectively. In Fig.~\ref{Fig_HDMF}(b--c) we plot the mean log permeabilities for the smallest RMSE over simulations, which is 30.51 for ETPF and 32.48 for ETKF.
In Fig.~\ref{Fig_HDMF}(d--e) we plot the mean log permeabilities for the largest RMSE over simulations, which is 39.2 for ETPF and 33.87 for ETKF. 
We observe that ETKF as well as IS provide smooth mean permeability fields that have smaller absolute values than the true permeability. ETPF gives higher variations of the mean permeability field and is in an excellent agreement with the true permeability for a good initial ensemble shown in Fig.~\ref{Fig_HDMF}(b). 
This means that ETPF sensitivity to the initial sample is due to sampling error and that the spatial variability of ETPF is a result of  sampling error.
It should be noted that IS with ensemble size $10^3$ and this good initial ensemble gives the RMSE 30.51 and the same mean log permeability field as ETPF shown in Fig.~\ref{Fig_HDMF}(b). However, IS does not change the parameters, only their weights, while ETPF does change the parameters. Therefore ETPF has an advantage of IS representing the correct posterior but does not have its disadvantage of resampling lacking.
In Fig.~\ref{Fig_HDVF} we plot the variance of the permeability fields obtained with ensemble size $10^5$ by IS (d), with ensemble size $10^3$ by ETPF (b--e) and ETKF (c--f). Fig.~\ref{Fig_HDVF}(b--c) is for the smallest RMSE and Fig.~\ref{Fig_HDVF}(e--f) is for the largest RMSE. 
ETKF provides smoother variance than ETPF due to smaller sampling errors. 

\begin{figure} [t]
	\centering		
	\includegraphics[width=1.0\linewidth]{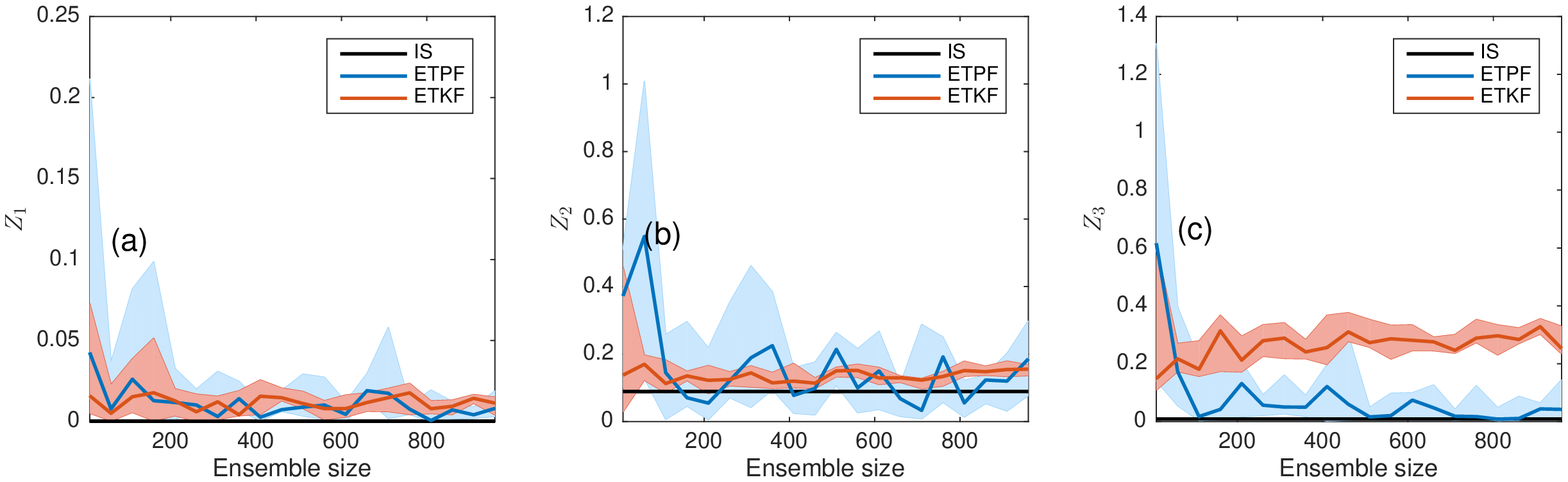}	
	\caption{Squared error between the true and the mean estimated modes for $\mathcal{Z}_1$ (a), $\mathcal{Z}_2$ (b), and $\mathcal{Z}_3$ (c) w.r.t ensemble size. ETPF is shown in blue and ETKF in red with solid lines for median and shaded area for 25 and 75 percentile over 10 simulations. IS with ensemble size $10^5$ is in black.}	\label{Fig_HDpar}
\end{figure}

\begin{figure} [t]
	\centering		
	\includegraphics[width=1.0\linewidth]{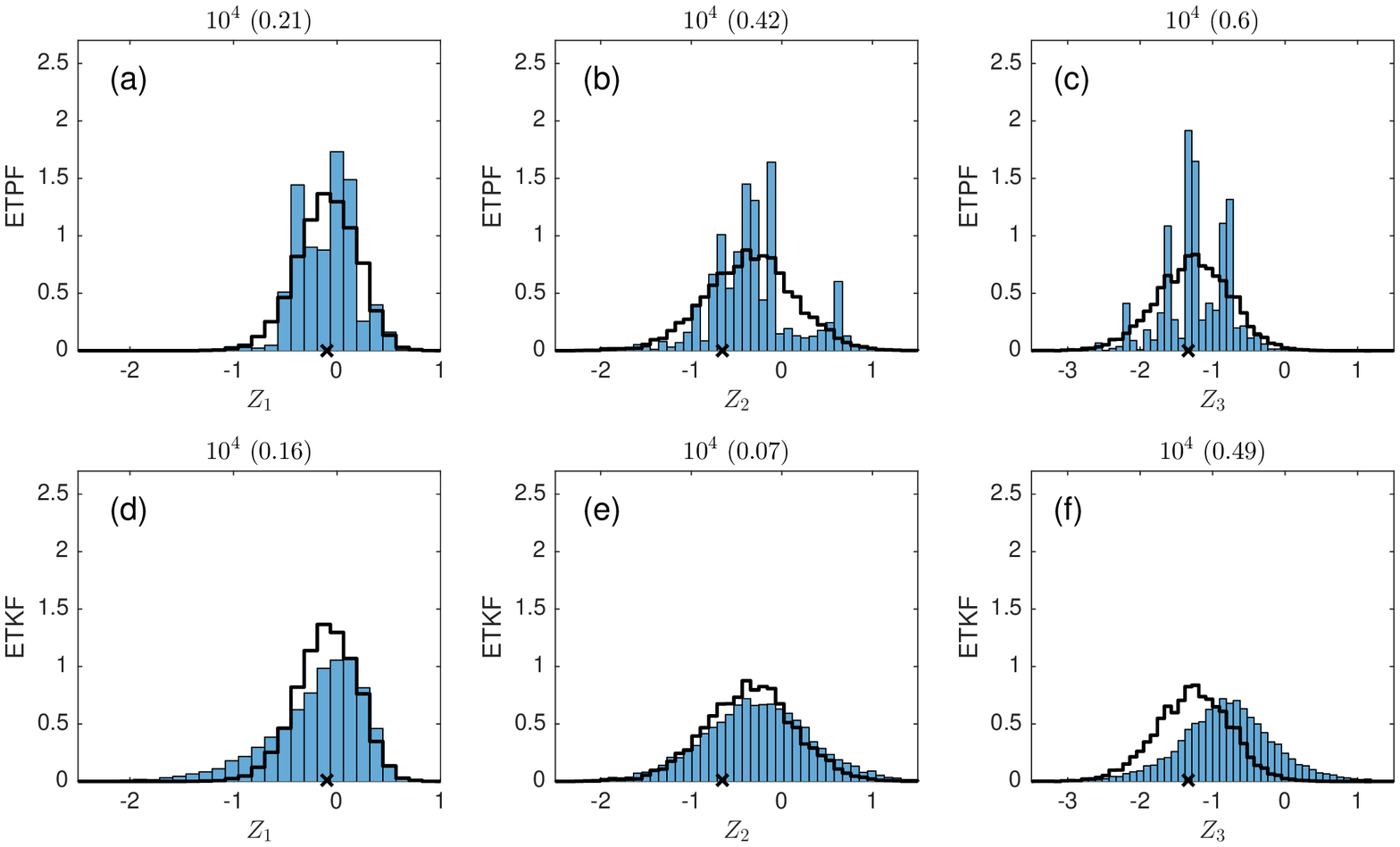}	
	\caption{The posterior probability density function of parameters  $\mathcal{Z}_1$ (left),  $\mathcal{Z}_2$ (center), and $\mathcal{Z}_3$ (right). The posterior obtained by IS with ensemble size $10^6$ is plotted as a black line and the true parameter as a black cross. The posterior of ETPF is shown at the top and the posterior of ETKF at the bottom. Both ETPF and ETKF used $10^4$ ensemble members. The Kullback-Leibler divergence is in brackets.}	
	\label{Fig_HDpdf}
\end{figure}

\begin{figure} [t]
	\centering		
	\includegraphics[width=1.0\linewidth]{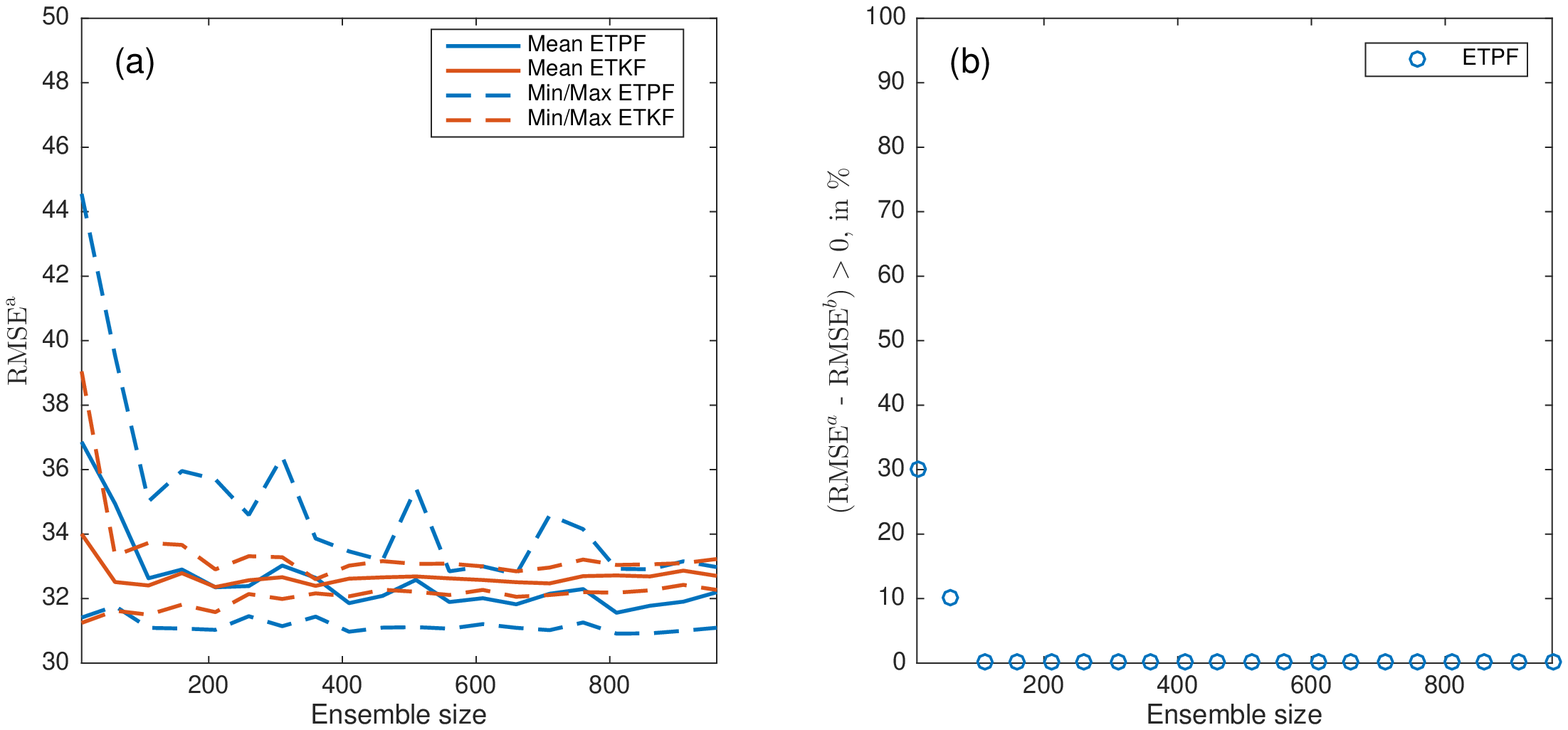}	
	\caption{Using only three leading modes in the KL expansion. Panel (a): RMSE after data assimilation w.r.t ensemble size with mean, minimum and maximum over 10 simulations for ETPF shown in blue and ETKF in red. Panel (b): $\%$ of simulations that result in $(\text{RMSE}^{a}-\text{RMSE}^{b})>0$ for ETPF.}	
	\label{Fig_HDLMerr}
\end{figure}

\begin{figure} [t]
	\centering		
	\includegraphics[width=1.0\linewidth]{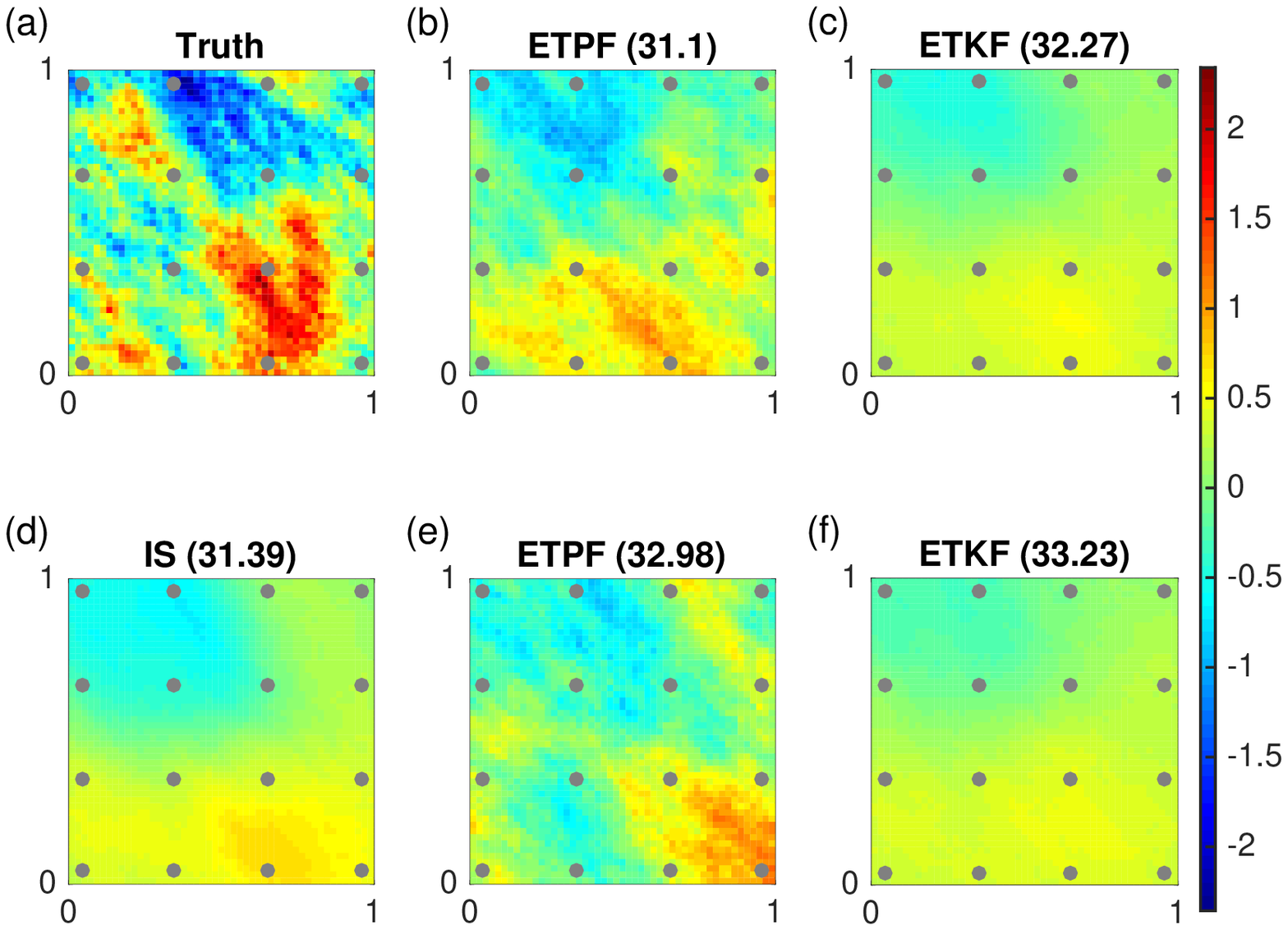}	
	\caption{Same as figure~\ref{Fig_HDMF}, but using only three leading modes in the KL expansion.}	\label{Fig_HDLMMF}
\end{figure}

In Fig.~\ref{Fig_HDpar} we show squared error $(\overline{\mathcal{Z}}^a - \mathcal{Z}^\textrm{true})^2$ 
in blue for ETPF and in red for ETKF for three leading modes $\mathcal{Z}_1$ (a), $\mathcal{Z}_2$ (b), and $\mathcal{Z}_3$ (c), 
where solid line is for median and shaded area is for 25 and 75 percentile over 10 simulations. 
We observe that in terms of the estimation of the three leading modes ETPF outperforms ETKF. 
In Fig.~\ref{Fig_HDpdf} we plot the posterior of $\mathcal{Z}_1$ (left),  $\mathcal{Z}_2$ (center), and $\mathcal{Z}_3$ (right) obtained by 
IS with ensemble size $10^6$ and by ETPF (top) and ETKF (bottom) with ensemble size $10^4$. 
The posterior of these modes is roughly approximated by ETPF as shown in Fig.~\ref{Fig_HDpdf} (a)--(c). 
ETKF provides a skewed posterior of the modes shown in Fig.~\ref{Fig_HDpdf} (d)--(f), which was also observed in the one parameter nonlinear problem, 
see Fig.~\ref{Fig_1Dpdf}(f). In order to perform an objective comparison between the probabilities we compute the Kullback-Leibler divergence of a 
posterior $\pi$ obtained by either ETPF or ETKF and the posterior $\pi^\textrm{IS}$ obtained by IS according to Eq.~\eqref{DKL}.
ETPF gives the Kullback-Leibler divergence 0.21, 0.42, and 0.6, while ETKF 0.16, 0.07, and 0.5 for the
modes $\mathcal{Z}_1$, $\mathcal{Z}_2$, and $\mathcal{Z}_3$, respectively. Thus ETKF gives a better approximation of the true pdf. 
Since first modes are well estimated by ETPF and last modes are not (not shown), 
we use only three leading modes in the Karhunen-Loeve expansion given by Eq.~(\ref{KLexp}) when computing the estimated log permeability 
keeping the number of uncertain parameters the same, namely 2500. In Fig.~\ref{Fig_HDLMerr}(a) we observe that ETPF outperforms ETKF for large ensemble sizes independent of an initial sample. Moreover, ETPF is not overfitting the data anymore since RMSE always decreases after data assimilation except at small ensemble sizes shown in Fig.~\ref{Fig_HDLMerr}(b). In Fig.~\ref{Fig_HDLMMF} we show the mean fields for the best and worst initial samples of $10^4$ size. ETPF gives RMSE at the best sample 31.1 and  the worst sample 32.98. By comparing it to 30.51 and 39.2 obtained using the full Karhunen-Loeve expansions, we observe that the maximum RMSE over simulations decreased substantially, while the minimum RMSE only slightly increased.   
ETKF gives RMSE at the best sample 32.27 and the worst sample 33.23. (Compare to 32.48 and 33.9 using the full Karhunen-Loeve expansions). 
Thus ETKF slightly decreases both maximum and minimum RMSE over simulations. 
ETPF is more affected by sampling noise at small scales, 
so using a truncated representation of the fields significantly improves the results for ETPF. 
ETKF is filtering out the small scales that are not observed and thus is less affected by the truncation.

\begin{figure} [t]
	\centering		
	\includegraphics[width=1.0\linewidth]{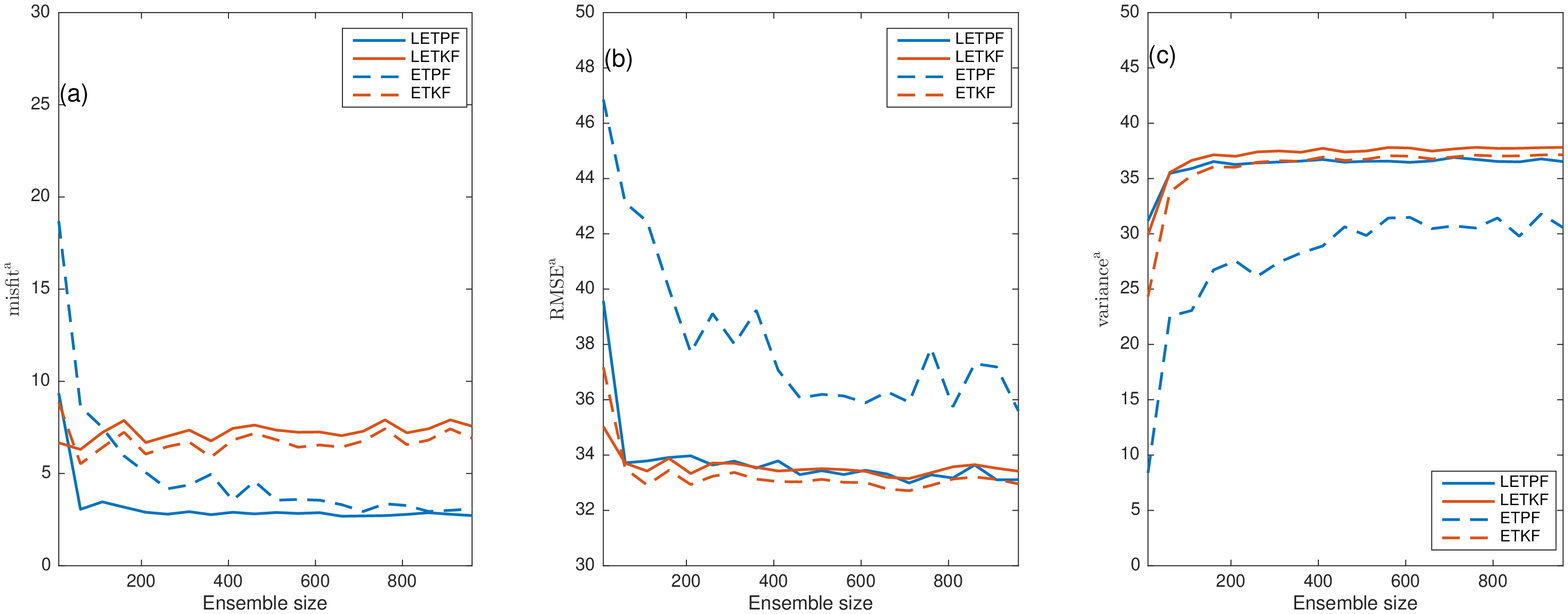}	
	\caption{Mean over 10 simulations after data assimilation for the data misfit (a), RMSE (b), and variance (c). 
		LETPF is shown in solid blue and LETKF in solid red.
		ETPF is shown in dashed blue and ETKF in dashed red.}	
	\label{Fig_HDLocerr}
\end{figure}

Next we apply LETPF and LETKF. The optimal localization radius between 0.2 and 1.2 was obtained in terms of the smallest RMSE and shown in Table~\ref{Tab_rho}.
It should be noted that smaller localization radius for LETPF than for LETKF was also observed by~\cite{ChRe15} 
and it is probably related to more noisy approximation of the posterior by LETPF than by LETKF. In Fig.~\ref{Fig_HDLocerr} we plot misfit, RMSE and variance. 

\begin{table}[t]
		\centering
	\begin{center}
	\caption{Optimal localization radius for LETPF and LETKF at different ensemble sizes M.\label{Tab_rho}}
	\begin{tabular}{lccccr}
		
        \hline 
		M & 10  &  110  &  210  &  $\hdots$ & 910 \\
        \hline \\
		LETPF & 0.2  &  0.6    &   0.6  &  $\hdots$ & 0.6\\
		\\
		LETKF & 0.2  &  1.2    &   1.2  &  $\hdots$ & 1.2\\
         \hline
	\end{tabular}
	\end{center}
\end{table}
%
At small ensemble sizes both LETKF and LETPF give smaller misfit, smaller RMSE but larger variance than ETKF and ETPF.
For large ensembles LETKF performs worse than ETKF, which is due to the imposed range 
on localization radius, meaning that 1.2 is not optimal.  
Comparing the performance of LETPF to (L)ETKF we observe 
that at small ensemble sizes LETKF still outperforms ETPF
but at large ensemble sizes LETPF performs now comparably to ETKF. 
Moreover, LETPF overfits the data less often than ETPF: $40\%$ against $90\%$ for ensemble size 10 
and $0\%$ against non-zero$\%$ for ensemble sizes greater than 150 (not shown). 

\begin{figure} [t]
	\centering		
	\includegraphics[width=1.0\linewidth]{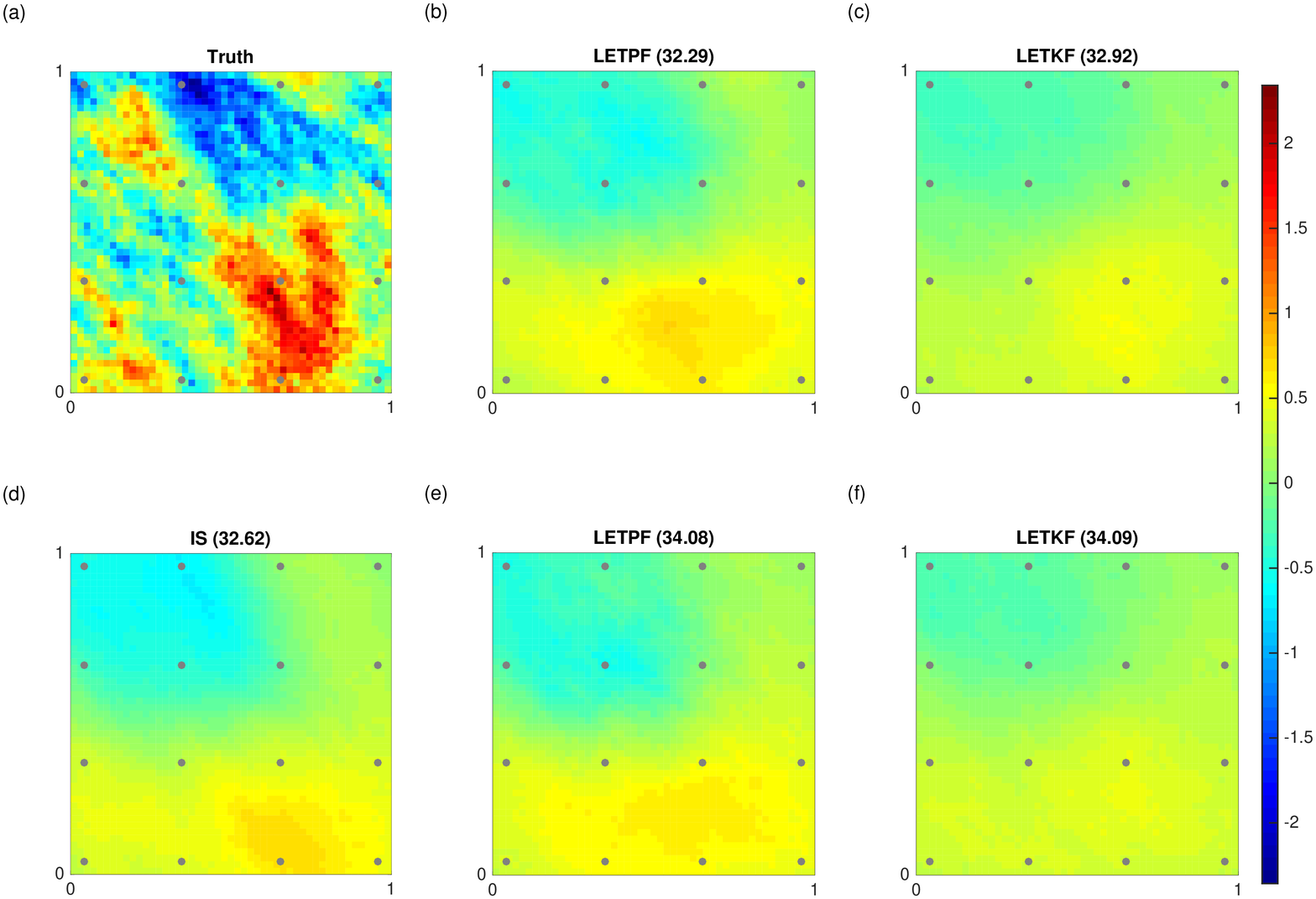}	
	\caption{Same as figure~\ref{Fig_HDMF}, but with localization.}	
	\label{Fig_HDLocMF}
\end{figure}

\begin{figure} [ht]
	\centering		
	\includegraphics[width=1.0\linewidth]{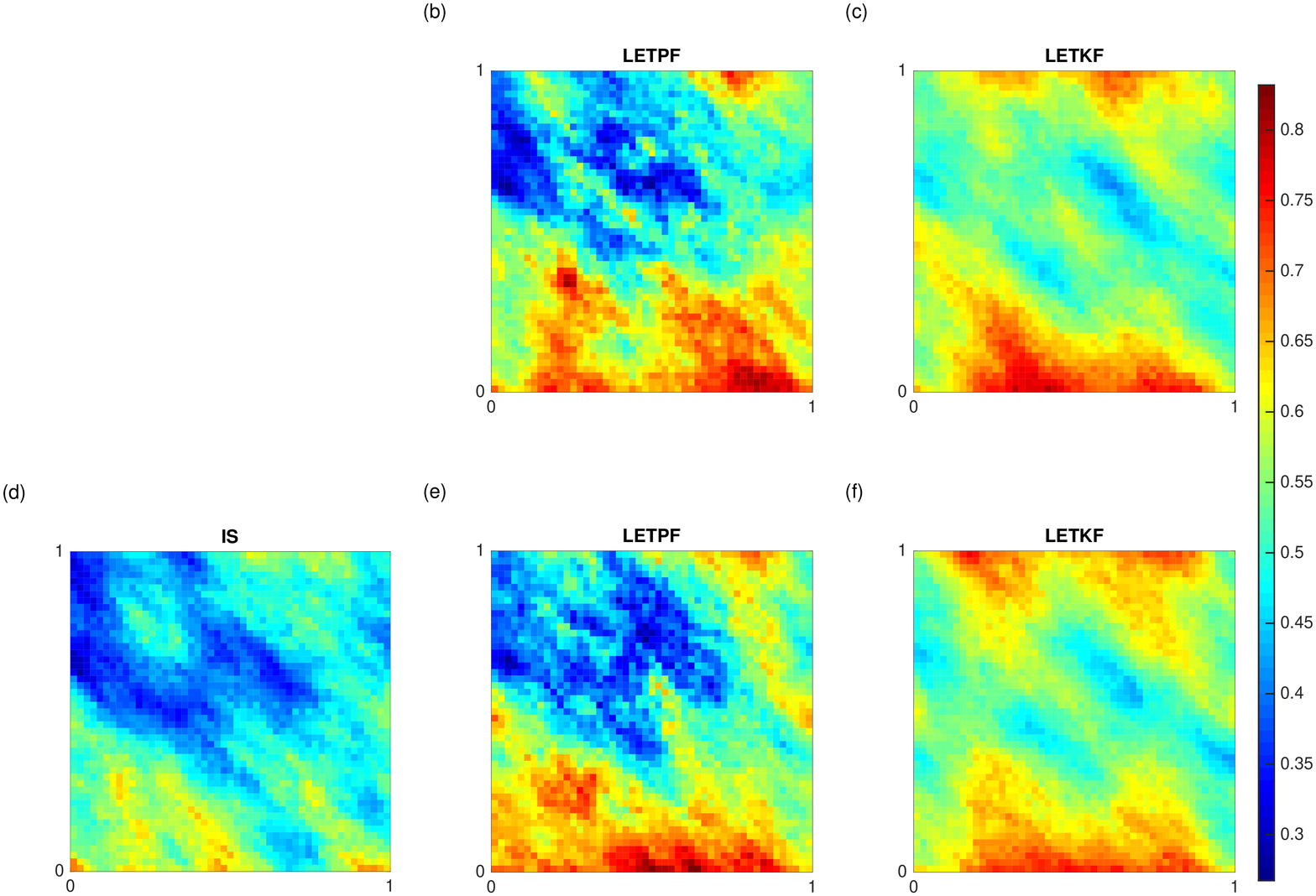}	
	\caption{Same as figure~\ref{Fig_HDVF}, but with localization.}	
	\label{Fig_HDLocVF}
\end{figure}

In Fig.~\ref{Fig_HDLocMF}--\ref{Fig_HDLocVF} we plot mean and variance of the log permeability field at ensemble size $10^3$ for ETPF (b)--(e) and ETKF (c)--(f) with localization at the smallest RMSE (b)--(c) and largest RMSE (e)--(f) over simulations, which are 32.29 and 34.08 for ETPF and 32.92 and 34.09 for ETKF, respectively. We observe that localization decreases the sampling noise and the spatial variability of the mean field obtained by ETPF at ensemble size $10^3$ resembles IS at ensemble size $10^5$. The variance obtained by ETPF with localization shown in Fig.~\ref{Fig_HDLocVF}(b--e) has also improved. 

The posterior estimation of the leading mode $\mathcal{Z}_1$, however, degraded, while of $\mathcal{Z}_2$ and $\mathcal{Z}_3$ improved. 
The Kullback-Leibler divergence for the leading mode is 0.73 (compare to 0.21 without localization), and for second and third is 0.2 and 0.18, 
correspondingly (compare to 0.42 and 0.6 without localization). Variance of the posteriors is larger when localization is applied for both methods. 
The localized weights given by Eq.~\eqref{PF_wgtR0} vary less than the non-localized weights given by Eq.~\eqref{PF_wgt}. 
Therefore the localized pdf is less noisy than the non-localized. 
However, localization applied in the form of the Karhunen-Loeve expansion given by Eq.~\eqref{KLexp} does not retain the imposed bounds on the modes $\mathcal{Z}$ 
as we need to invert a matrix product of eigenvalue and eigenvector matrices to obtain the modes. 
Moreover unlike ETKF, LETPF does not converge to ETPF as the localization radius 
goes to infinity due to the transport problem being univariate for LETPF and multivariate for ETPF. 


\section{Conclusions} \label{Sec:Con}  
MCMC methods remain the most reliable methods for estimating the posterior distributions of uncertain model parameters and states. They, however, also remain computationally expensive. Ensemble Kalman filters provide computationally affordable approximations but rely on the assumptions of Gaussian probabilities. For nonlinear models even if the prior is Gaussian the posterior is not Gaussian anymore. Particle filtering on the other hand does not have such an assumption but requires a resampling step, which is usually stochastic. 
Ensemble transform particle filter is a particle filtering method that deterministically resamples the particles based on their importance weights and covariance maximization among the particles.

ETPF certainly outperforms ETKF for a one parameter nonlinear test case by giving a better posterior estimation. This conclusion also holds for the five parameter test case, however demands a substantially larger ensemble size. Moreover the mean estimations obtained by ETPF are not consistently better than the ones obtained by ETKF. 
When the number of uncertain parameters is large (2500) a decrease of degrees of freedom is essential. This is performed by  localization. At large ensemble sizes ETPF performs as well as ETKF, while at small ensemble sizes ETKF still outperforms ETPF.
Even though localized ETPF overfits the data less often than non-localized, localization destroys the property of ETPF to retain the imposed bounds. This results in deterioration of the first mode posterior approximation. Another approach to improve ETPF performance is instead of applying localization to use only first modes in the approximation of log permeabilty as they are better estimated by the method. An advantage of this approach is that it is fully Bayesian. However, one needs to know at which mode to make a truncation and this is highly dependent on the covariance matrix of the log permeability.

\section*{Acknowledgments} 
This work is part of the research programme Shell-NWO/FOM Computational Sciences for Energy Research (CSER) with project number 14CSER007 which is partly financed by the Netherlands Organization for Scientific Research (NWO).

\bibliographystyle{plain}
\bibliography{RuDu18.bib}

	\end{document}